\documentclass[twocolumn,prm, letterpaper,showpacs,preprintnumbers,amsmath,amssymb]{revtex4-1}

\usepackage{graphicx}
\usepackage{dcolumn}
\usepackage{bm}
\usepackage{epsfig}

\begin{document}

\title{Topological surface states of MnBi$_2$Te$_4$ at finite temperatures and at domain walls}

\author{Kevin F. Garrity}
\email{kevin.garrity@nist.gov}
\affiliation{%
Material Measurement Laboratory, National Institute of Standards and Technology, Gaithersburg MD, 20899
}%

\author{Sugata Chowdhury}
\affiliation{%
Material Measurement Laboratory, National Institute of Standards and Technology, Gaithersburg MD, 20899
}%

\author{Francesca M. Tavazza}
\affiliation{%
Material Measurement Laboratory, National Institute of Standards and Technology, Gaithersburg MD, 20899
}%

\date{\today}

\begin{abstract}
MnBi$_2$Te$_4$ has recently been the subject of intensive study, due
to the prediction of axion insulator, Weyl semimetal, and quantum
anomalous Hall insulator phases, depending on the structure and
magnetic ordering. Experimental results have confirmed some aspects of
this picture, but several experiments have seen zero-gap 
surfaces states at low temperature, in conflict with expectations. In
this work, we develop a first-principles-based tight-binding model
that allows for arbitrary control of the local spin direction and
spin-orbit coupling, enabling us to accurately treat large
unit-cells. Using this model, we examine the behavior of the
topological surface state as a function of temperature, finding a gap
closure only above the N\'eel temperature. In addition, we examine the effect of
magnetic domains on the electronic structure, and we find that the domain
wall zero-gap states extend over many unit-cells. These domain wall states can appear
similar to the high temperature topological surface state when many domain sizes
are averaged, potentially reconciling theoretical results with
experiments.
\end{abstract}

\maketitle

\section{Introduction}

Since the pioneering work of Haldane\cite{haldane}, there has been
great interest in the topological properties of materials systems,
with many exciting developments in the past dozen
years\cite{ti_review, ti_review2, tokura2019magnetic}. However, much
of the progress on topological systems has been focused on topological
classes with time-reversal symmetry (TRS), and topological materials
with broken TRS, \textit{i.e.} magnetic materials, remain challenging
to design and study. The zero-field quantum anomalous Hall effect in
particular has only been realized in magnetically-doped topological
insulators, with sub-Kelvin temperatures necessary to observe robust
quantization, limiting possible applications of this
effect\cite{chen2010massive, chang2013experimental, chang2015high, lee2015imaging}.


MnBi$_2$Te$_4$ and MnBi$_2$Se$_4$ have recently been the subject of
intensive study\cite{lee2013crystal, hagmann2017molecular,
  deng2019magneticfieldinduced, liu2020robust, chen2019intrinsic,
  PhysRevB.100.121104, swatek2019gapless, PhysRevX.9.041039,
  PhysRevX.9.041038, PhysRevX.9.041040, PhysRevResearch.1.012011,
  xiao2018realization,li2019intrinsic, gong2019experimental,
  Rienks2019}, due to theoretical predictions\cite{Chowdhury2019,
  zhang2019topological, Otrokov2019, eremeev2018new,
  otrokov2017magnetic, otrokov2017highly, higherorder} that they are
antiferromagnetic (AFM) topological insulators (TI), a type of axion
insulator, in bulk\cite{mong2010antiferromagnetic, li2010dynamical,
  wang2016dynamical, essin2009magnetoelectric}. In addition, they can
display Weyl semimetal phases under strain and/or external magnetic
field. In two-dimensional geometries, they are predicted to be Chern
insulators for systems with an odd number of layers. This materials
class offers the possibility of observing broken-TRS topological
effects in single-crystal materials with reasonably high magnetic
transition temperatures ($T_N \approx 24$~K\cite{Otrokov2019}) and
large band gaps, which should improve the robustness of the
topological effects. However, there has been some disagreement between
experiments and theoretical expectations, and in some cases between
different experiments, on fundamental properties of this
material. Under external magnetic field sufficient drive a transition
from the layered AFM ground state to a fully spin-polarized
ferromagnetic (FM) state, the quantum anomalous Hall effect has been
observed, as expected, but the anomalous Hall conductivity (AHC) of
odd-layer systems is not observed to be quantized at zero
field\cite{deng2019magneticfieldinduced, liu2020robust}. In addition,
several experiments have observed surface state features even below
the N\'eel temperature\cite{PhysRevX.9.041039, PhysRevX.9.041038,
  PhysRevX.9.041040, swatek2019gapless, Otrokov2019,
  PhysRevB.100.121104, vidal2019topological, PhysRevB.101.161109, PhysRevLett.125.117205}, which are expected to be
gapped by the broken TRS on the surface, while other experiments have
seen inconsistent or conflicting
results\cite{PhysRevResearch.1.012011,chen2019intrinsic,
  PhysRevB.100.121104, zeugner2019chemical, chang2013experimental}
(see discussion in Ref.~\onlinecite{PhysRevX.9.041040}). Either a
different surface magnetic ordering or domain
walls\cite{PhysRevX.9.041038, PhysRevX.9.041039, PhysRevX.9.041040, PhysRevB.101.161109}
have been suggested as possible explanations for the low temperature
surface states. Recently, the presence of domain walls has been
confirmed using atomic-force microscopy\cite{domainwall}, but at
relatively low densities, while robust surface A-type AFM ordering has
also been confirmed\cite{atype, PhysRevLett.125.117205}.

To address these discrepancies, in this work, we develop a first
principles-based model of the magnetic degrees of freedom and
electronic structure of MnBi$_2$Te$_4$ that can be applied to large
unit cells. Using this model, we can directly calculate some of the
proposed scenarios for explaining the various experimental results,
which may help clarify the experimental situation. We first briefly
consider the iso-symmetric topological transition the occurs as a
function of spin-orbit coupling (SOC) strength. Next, we study the
temperature-driven topological phase transition that accompanies the
N\'eel transition, observing how the bulk and surface band structures
change in response to changes in the spin ordering. We find that
consistent with expectations, the system has a bulk band gap both
above and below the transition temperature, but only has a surface
state above $T_N$ when TRS is restored. We also consider
configurations with partially ordered surface spins, and we find that
such configurations can cause the surface gap to close if the disorder
is large enough, even if the surface still has broken TRS on
average. Finally, we study domain walls in low temperature
MnBi$_2$Te$_4$, which can be understood as a type of topological
transition that occurs as a function of spatial
location\cite{mong2010antiferromagnetic,domain_theory,
  Otrokov2019}. We find spin-polarized metallic edge states localized
on the surface at the domain walls, but that extend over many
unit-cells along the surface perpendicular to the domain wall. These
surface features can appear similar to the topological surface states
we observe in the disordered spin configurations, which may help
reconcile some of the unexpected experimental observations with
theory.

\begin{figure}
\includegraphics[width=3.6in]{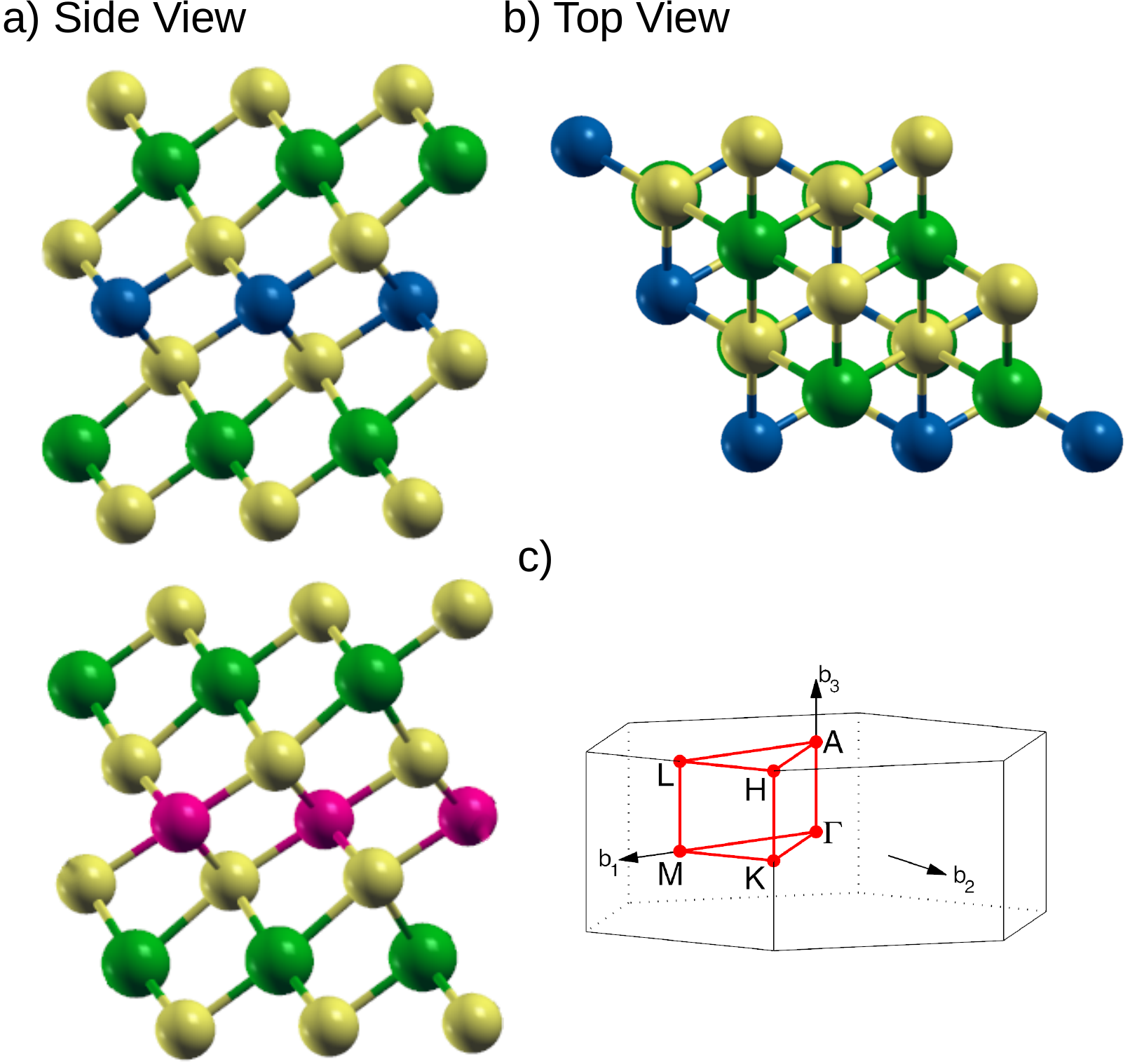}
\caption{\label{fig:crys} a-b) Side and top view of layered AFM structure of MnBi$_2$Te$_4$. Mn spin up is red, spin down is blue, Te is yellow and Bi is green. c) Brillioun zone. We use hexagonal labels to more easily compare bulk and surface calculations.}
\end{figure}

We show the crystal structure of MnBi$_2$Te$_4$, space group
$R\bar{3}m$, in Fig.~\ref{fig:crys}. The structure consists of a stack of seven
atom layers (septuple layers). In the ground state, the Mn within each
layer are ordered ferromagnetically, and alternating layers are
aligned antiferromagneticaly, with spins oriented along the $\pm z$
direction.

\section{Methods}

\subsection{First principles}

We perform first principles density functional theory (DFT)
calculations\cite{hk, ks} with the Quantum Espresso code\cite{qe}
using the PBEsol\cite{pbesol} functional. We use a DFT+U correction
with $U{=}3$ eV on the Mn-$d$ states\cite{ldaplusU, ldaplusU_simplified,
  dftU_values}. We use norm-conserving ONCV pseudopotentials with
SOC\cite{oncv, oncv_opt, oncv_so}. We use
Wannier90 \cite{wannier90, mlwf, mlwf_orig, mlwf_entangle} to
generate first principles tight-binding Hamiltonians, and we
calculate topological invariants with both
WannierTools\cite{wanniertools} and our own code. Our initial Wannier
projection consists of Bi/Te-$p$ orbitals and Mn-$d$ orbitals, which
describes all the bands near the Fermi level. 

\subsection{Magnetic tight-binding model}

In order to calculate the electronic structure of the large unit-cells
that are necessary to treat structures with disordered spins or domain
walls, we develop a tight-binding model, based on Wannier
Hamiltonians, that allows us to calculate the electronic structure for
arbitrary orientations of the Mn spins, as well as variable SOC. The model is similar in spirit to the model in
Ref. \onlinecite{liu2013topological}, which treats chemical disorder
in topological insulators. The basis of our model is three separate
DFT plus Wannier calculations. First, we perform a calculation with
TRS and without SOC, getting the Hamiltonian $H^{TRS}$. Second, we
perform a calculation with TRS and SOC, getting $H^{TRS}_{SOC}$. By
subtracting these two Hamiltonians, we can isolate the SOC
contribution, $H_{SOC} = H_{SOC}^{TRS} - H^{TRS}$. Finally, we perform
a FM calculation without SOC, which is separated into independent spin
up ($H^{up}$) and spin down ($H^{dn}$) terms.

We then assemble the total model for a single unit-cell, $H_{tot}$:
\begin{eqnarray}\label{eq:cluster}
H_{avg} &=& \frac{1}{2} (H^{up} + H^{dn}) \\
H_{diff} &=& \frac{1}{2} (H^{up} - H^{dn}) \\
H_{tot} &=& H_{avg} \sigma_0 + H_{diff} ({\bf m \cdot \sigma}) + H_{SOC}, 
\end{eqnarray}
where the vector ${\bf m}$ is the normalized magnetization direction,
$\sigma_0$ is the identity matrix, and ${\bf \sigma}$ are the three
Pauli matrices. To generate tight-binding Hamiltonians for supercells
with different magnetic orderings (${\bf m}$'s) in each cell, we keep
the onsite terms as above and average the inter-cell matrix
elements. This approximation allows us to treat arbitrary magnetic
orderings based solely on FM DFT calculations, and we verify its
accuracy below.

To construct surfaces, we create supercells of the desired
thickness, but then set to zero any hoppings that would go across the
surface.  This approximation is reasonable for MnBi$_2$Te$_4$ because
of the layered crystal structure, and direct surface calculations show
that the surface relaxation energy of MnBi$_2$Te$_4$ is only 5
meV per surface unit cell. We can also artificially adjust the
magnitude of the SOC by multiplying the final term in the model by a
number between zero and one.

\begin{figure}
\includegraphics[width=3.3in]{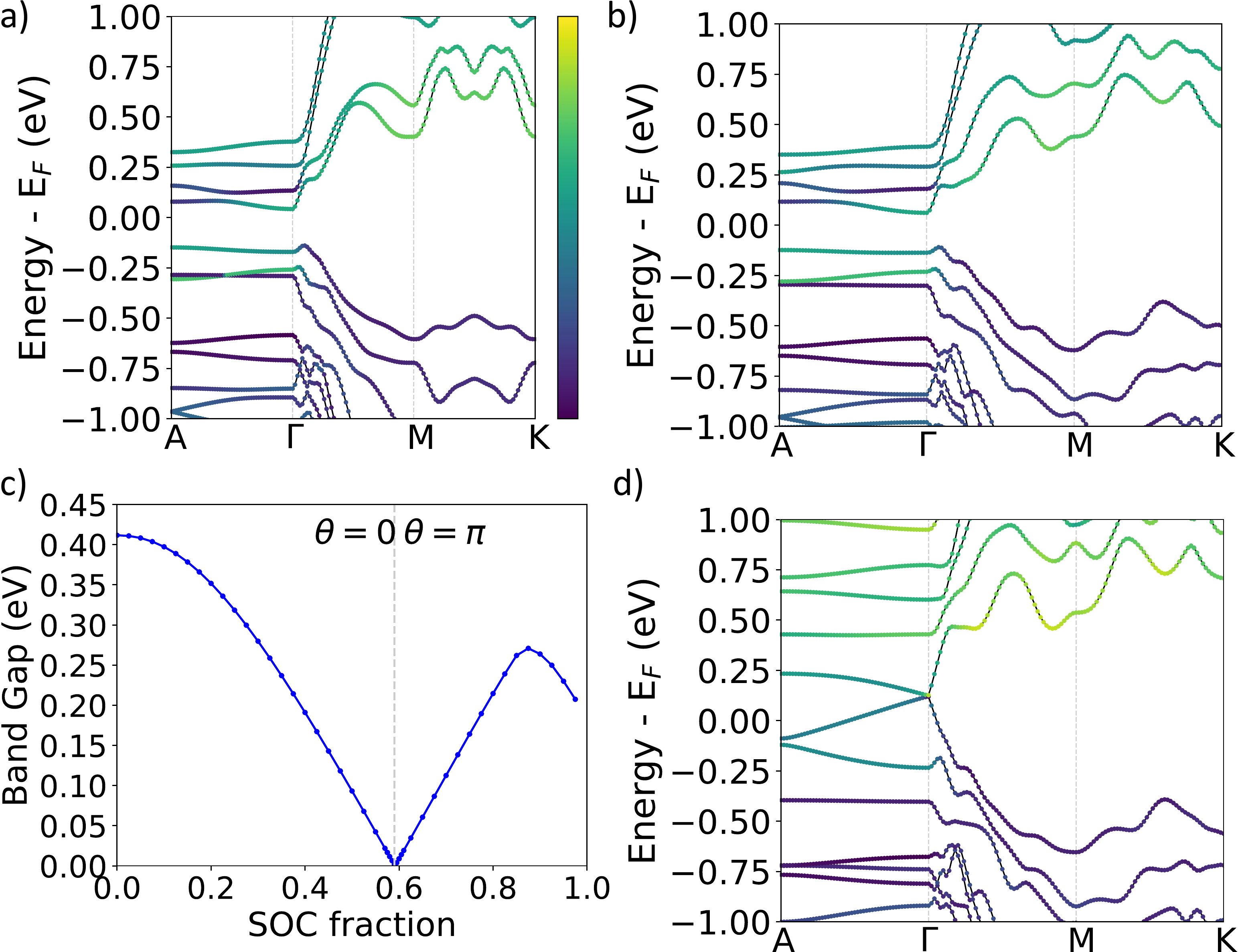}
\caption{\label{fig:soc} a) Band structure of AFM phase with spins in
  $z$-direction calculated using DFT. b) Same, but calculated with
  model. The colors show projections onto Bi Wannier functions. c)
  Band gap in eV as a function of SOC fraction. d) Model band
  structure at SOC=0.59, the critical value.}
\end{figure}

In order to verify the accuracy of this model, we compare the model
band structure to equivalent calculations done directly with
DFT-derived Wannier Hamiltonians for several spin configurations. In
Fig.~\ref{fig:soc}a and b, we show the DFT and model band structures
for the ground state AFM phase with spins in the $\pm
z$-direction. Comparing the two figures, we find excellent agreement,
with all major features of the band structure reproduced by the
model. We emphasize that the model is built using only FM spin
configurations and only non-magnetic SOC calculations, so its success
describing an AFM calculation with SOC is encouraging. We show several
more bulk band structure comparisons with various spin orderings in
Fig.~S1 of the supplementary materials (SM)\cite{supmat}. In
Fig.~\ref{fig:surface}, we directly compare a three-layer thick
surface DFT calculation with our model. For both the bulk and surface
calculations, we find excellent agreement. In order to interpret the
band structures of systems with large unit cells and magnetic
disorder, we use band unfolding to produce effective primitive cell
spectral functions\cite{liu2013topological, ku2010unfolding,
  berlijn2011can, popescu2010effective}.

\begin{figure}
\includegraphics[width=3.3in]{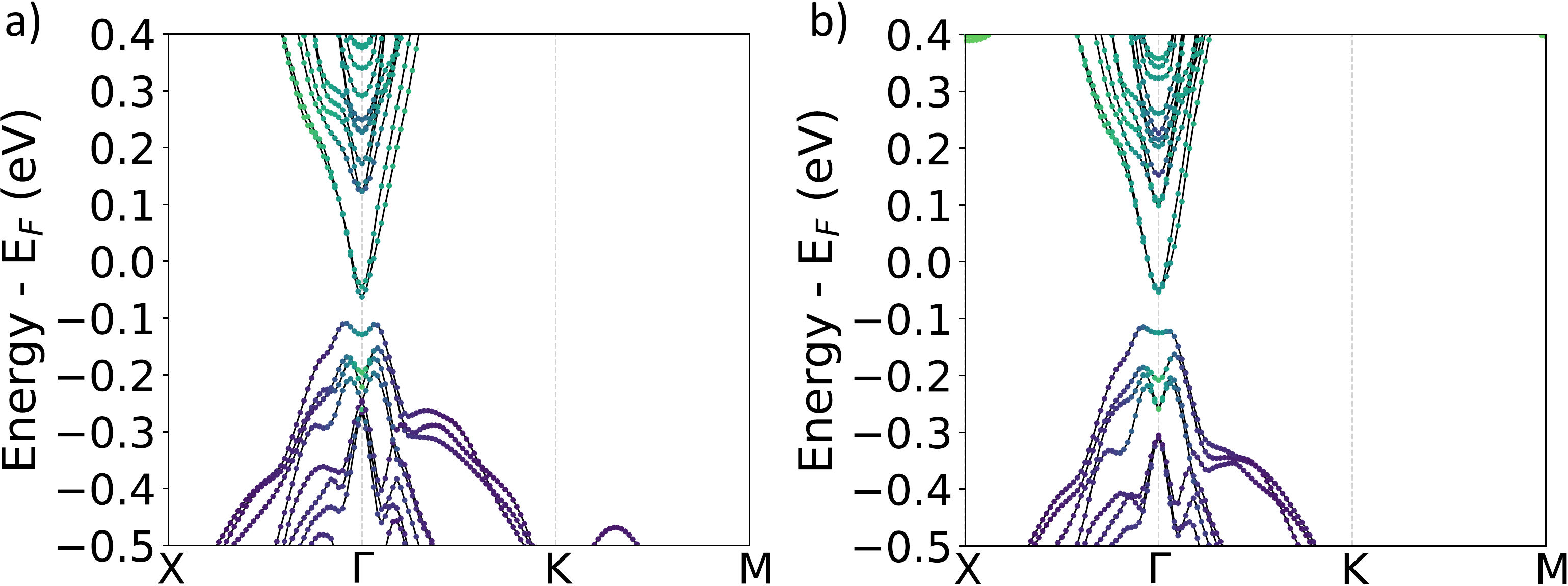}
\caption{\label{fig:surface} a) DFT and b) model band structure of three layer slab with surfaces, with out-of-plane AFM spin ordering. The DFT calculation includes surface relaxation. Colors as in Fig.~\ref{fig:soc}.}
\end{figure}

\subsection{Spin model \label{sec:spin}}

Similar to previous works on this material class\cite{Chowdhury2019,
  Otrokov2019}, we model the energetics of the spin-spin interactions
in our system using a Heisenberg model with onsite anisotropy.

\begin{eqnarray}\label{eq:heis}
H = \frac{1}{2} \sum_{ij} J_{ij} \vec{S_i} \cdot \vec{S_j} + \sum_i A |S^z_i|^2
\end{eqnarray}

We treat the $\vec{S_i}$ variables as classic spins. We fit the
coupling parameters, $J_{ij}$ and $A$, using a least squares approach
and taking into account crystal
symmetries\cite{cluster_spring}. Symmetry allows for additional
anisotropic intersite coupling terms in magnetic materials with
spin-orbit coupling; however, given the success of the previous works\cite{Otrokov2019}
describing magnetic interactions in MnBi$_2$Te$_4$, we limit our model
to Heisenberg intersite terms. We fit to DFT calculations of various
spin configurations in the equivalent of $2 \times 2 \times 2$ and $3
\times 3 \times 3$ unit cells, using the method of Lloyd-Williams
\textit{et al.} to generate smaller non-diagonal
cells\cite{nondiag}. We then use Metropolis Monte Carlo sampling to
generate spin configurations at a given temperature\cite{montecarlo}.

\section{Results}

\subsection{Variable spin-orbit}

Using our tight-binding model, we can now study changes in the
electronic structure during several types of topological phase
transitions. As a warm-up, we first consider the iso-symmetric
topological transition that occurs when artificially varying the
strength of the SOC. In Fig.~\ref{fig:soc}c, we show the band gap of
the ground state AFM phase as a function of the strength of
SOC. The non-trivial AFM topological state of MnBi$_2$Te$_4$ is driven by SOC-induced band inversion.  Therefore, at
zero SOC, MnBi$_2$Te$_4$ is a trivial AFM insulator. As the fraction
of SOC is increased, the bulk band gap closes, and at the critical
value of the SOC, 0.59, the the band structure becomes inverted. Above
this value, our model is in a topologically non-trivial AFM insulating
phase, which is also an axion insulator. This transition is an example
of an iso-symmetric transition between a topologically non-trivial and
trivial state, which requires a bulk gap closure. In practice, directly controlling the SOC
experimentally is not possible, but this transition might be similar
to a topological transition that occurs as a function of doping elements with weaker SOC into the MnBi$_2$Te$_4$ structure.

\subsection{Temperature-dependent electronic structure}

Next, we consider the topological phase transition that occurs as a
function of temperature. Above the N\'eel temperature, the spins in
MnBi$_2$Te$_4$ become disordered, restoring TRS on average and causing a topological phase transition.

We generate spin configurations at a given temperature using our
magnetic model (see Sec.~\ref{sec:spin}). We find that our model has a
transition temperature of 40~K, which is in reasonable agreement with
experiment, considering that quantum fluctuations lower transition
temperatures. As expected for a layered structure, we find that
within-layer spin-spin correlations are much larger than inter-layer
correlations, and remain small but non-zero above the transition
temperature (see Fig.~S2-S3 for more details).

\begin{figure}
\includegraphics[width=3.3in]{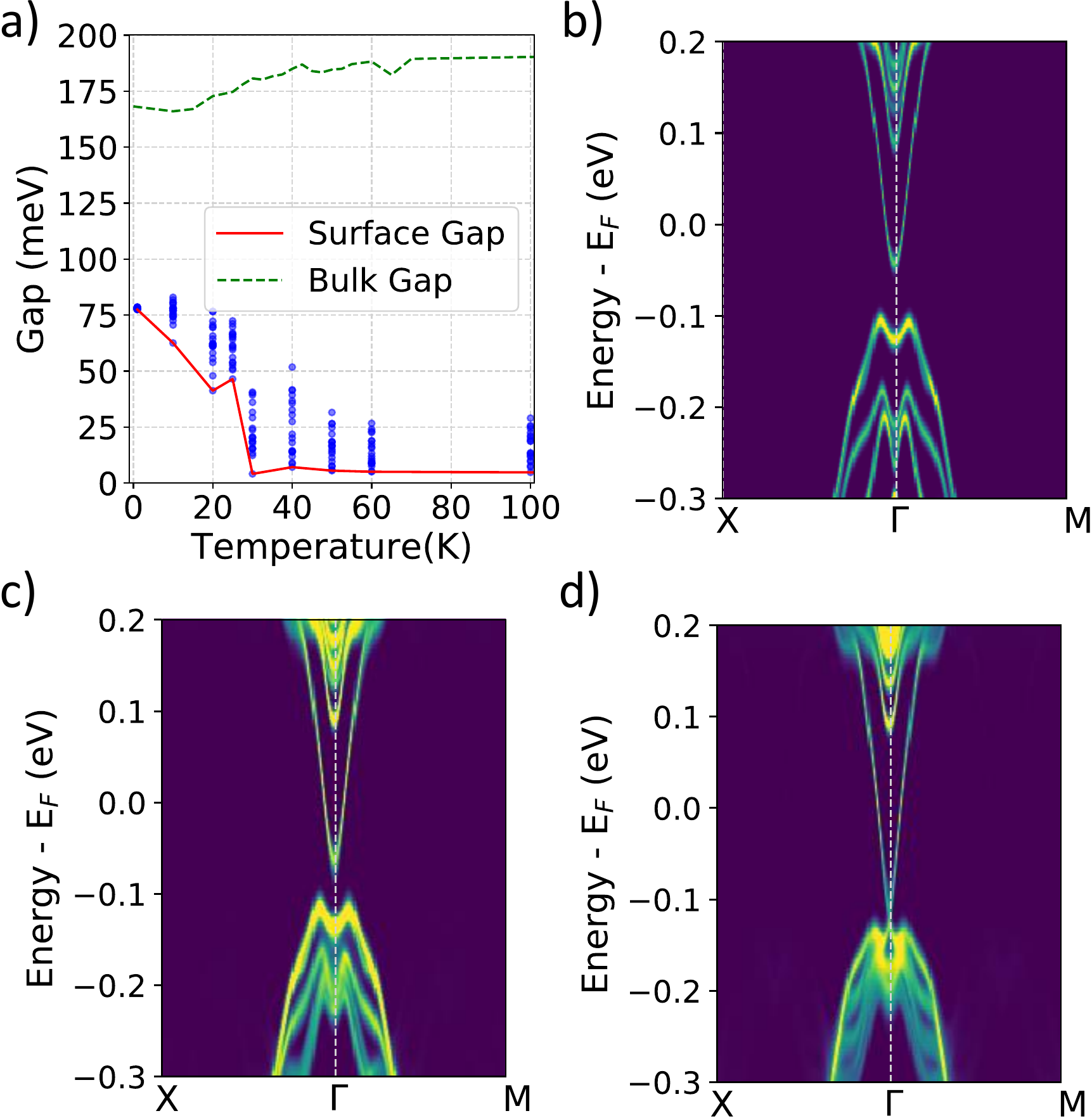}
\caption{\label{fig:temp} a) Band gap (meV) versus temperature. Solid
  red line: minimum surface gap. Dashed green line: mean bulk
  gap. Blue points: individual surface calculations. b-d) Unfolded
  average surface band structure at 1 K, 25 K and 100 K,
  respectively.}
\end{figure}

Using our magnetic model, we can generate spin configurations at a
given temperature, and then study the average electronic structure
using our tight-binding model. We first perform this analysis in a
periodic $3 \times 3 \times 6$ unit cell without a surface. We confirm
that the bulk gap does not close near $T_N$, which is consistent with
the fact that the bands near the Fermi level are primarily Bi and Te
states, with the Mn supplying spin-splitting. In
fact, the bulk gap opens slightly, as shown by the dashed line in
Fig.~\ref{fig:temp}a (see also supplementary materials
Fig. S5-S6)\cite{supmat}. Unlike the iso-symmetric SOC-driven transition studied
above, here, the order-to-disorder spin transition restores TRS at
high temperatures. Because of the symmetry change, the relevant
topological invariants are different above and below $T_N$, and no
bulk gap closure is required despite the topological transition.

Next, we monitor the same transition, but in $3 \times 3 \times 5$
unit-cell, with surfaces perpendicular to the $z$-direction. In this odd-layered
case, we find that MnBi$_2$Te$_4$ is a Chern insulator at zero
temperature, consistent with previous work\cite{Chowdhury2019, zhang2019topological, Otrokov2019}. In
Fig.~\ref{fig:temp}a, the blue points are the gaps of individual spin
configurations, and the solid red line shows the minimum gap at each
temperature. We find that above $T_N$, the minimum surface gap
closes. Individual spin configurations can have small gaps of
$\approx$25 meV even above $T_N$, which we attribute to spin
fluctuations breaking TRS. We expect that unit-cells with larger areas
than we can easily calculate would have smaller minimum gap fluctuations
above $T_N$, but that a spatially local measurement of the gap would continue to
fluctuate. 

In Figs.~\ref{fig:temp}b-d, we show the unfolded surface band
structure, averaged over 20 spin configurations, at $T=$1~K, 25~K, and
100~K, respectively. At low temperature, when the spins are almost
perfectly aligned, we find sharply defined bands and a clear band
gap. However, as the temperature is raised to 25 K, which is slightly
below the N\'eel temperature in our model, the bands become more
diffuse, and the spin-polarized bands begin to show the influence of
disorder. In addition, the gap at $\Gamma$ begins to close. Finally,
at 100 K, we find a closed gap, with a clear Dirac cone surface
feature, which shows that the system is in a non-trivial TRS-invariant ($Z_2\!=\!1$) topological
insulating state. This average topological state emerges despite the
fact that the individual band structures that go into the average
break TRS.

\subsection{Surface spin disorder}

\begin{figure}
\includegraphics[width=3.3in]{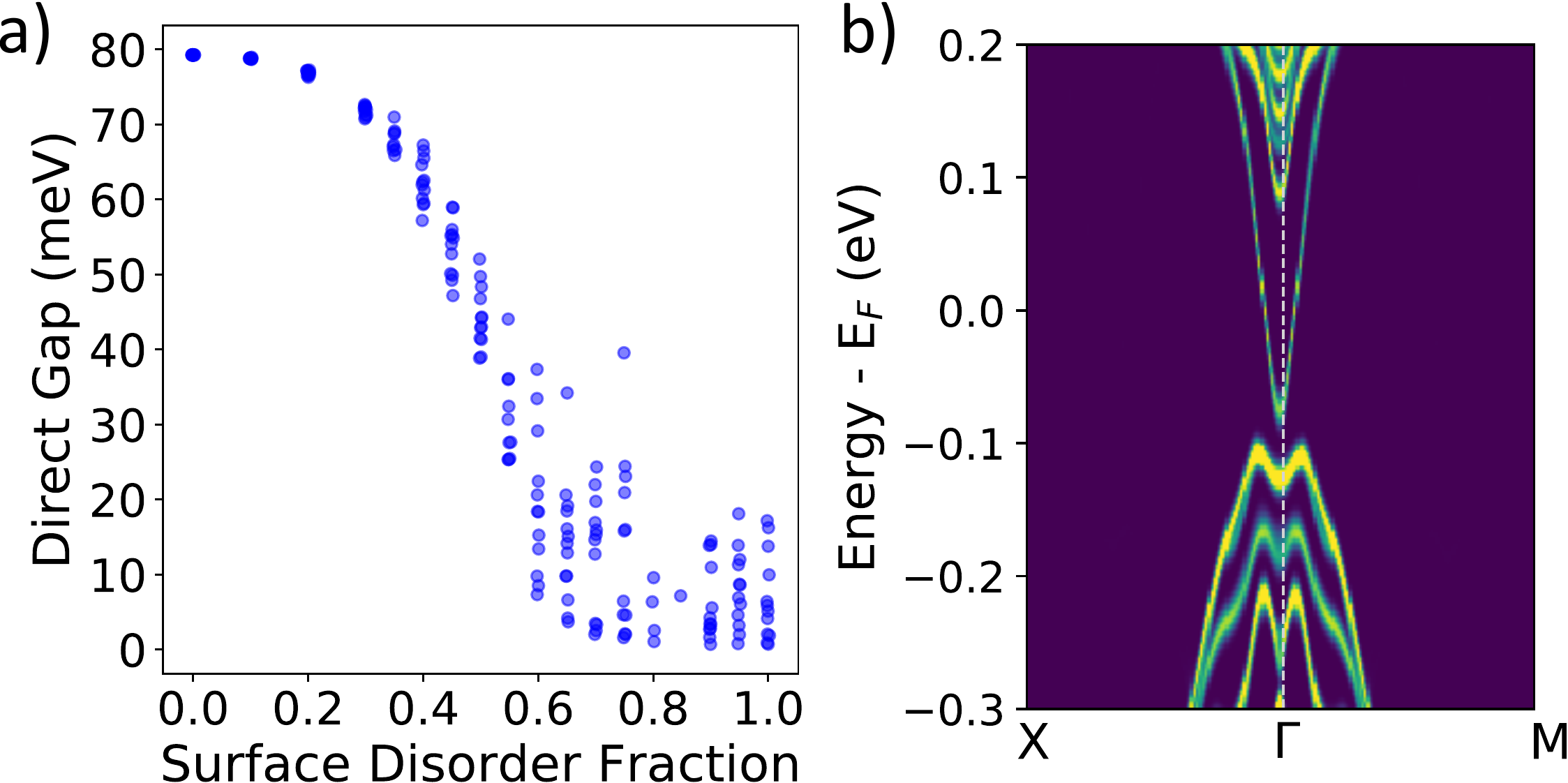}
\caption{\label{fig:disorder} a) Surface band gap (meV) of $3 \times 3 \times 5$ supercell with ordered bulk and partially disordered surface spins (see text). Each point is one spin configuration. b) Unfolded averaged band structure at 0.5 mixing between ordered and disordered surface spins.}
\end{figure}

To better quantify the amount of disorder necessary to close the
surface band gap, we again consider a $3 \times 3 \times 5$ supercell
with surfaces, but now we keep the bulk three layers fixed to a
perfectly ordered AFM configuration and consider partially disordered
surface spins. Specifically, we choose surface spins such that each
spin is a mixture of a perfectly ordered spin, oriented along the $z$
direction, and a randomly oriented spin. We consider a range of mixing
fractions from 0 (perfectly ordered surface) to 1 (fully
disordered). Notably, all of the surface spin configurations with
disorder fraction $<1$ have broken TRS on the surface on average. Despite this
broken TRS, we find that disorder fractions above 0.5 are enough to
close the average surface gap, as shown in
Fig.~\ref{fig:disorder}a,b. As discussed above, thermal flucuations
alone are not enough to close the surface band gap at temperatures
significantly below $T_N$. Other possible sources of spin disorder
include quantum spin fluctuations or chemical disorder.

\subsection{Domain wall electronic structure}

The above discussion of temperature-driven topological states provides
a clear explanation of the surface features observed experimentally
above $T_N$; however, the states observed at low temperature remain
unexplained. One possible explanation is that the low temperature
surface spin configuration does not match the theoretical
predictions. However, in this section, we consider the alternate
explanation that there is a significant density of domains in the AFM
phase at low temperature, possibly pinned by sample dependent
defects. In the bulk of an AFM topological insulator, the topological
index on either side of a domain boundary is the same, as the spin
configurations are related by a translation by one layer in the
$z$-direction. Therefore, a gap closure at the domain wall is not
required. Equivalently, the axion angle of both domains equals
$\pi \pm 2\pi$\cite{essin2009magnetoelectric, mong2010antiferromagnetic}. However, in the presence of a surface, this translation is no
longer possible.  Each surface of an AFM TI
contributes $\pm \frac{e^2}{2h}$ to the total AHC, with the sign
determined by the direction of the spins in the top layer\cite{mong2010antiferromagnetic}. Therefore,
there are two distinct topological phases at the surface of an AFM TI,
and a domain wall between these surfaces must have a 1D conducting
channel that contributes a total of $\pm \frac{e^2}{h}$ to the AHC. In
this work, we consider sharp Ising-like domain walls, where the spins
suddenly change from $+z$ to $-z$, or vice versa, at the boundary. Of
course, more complicated configurations where the spins rotate
gradually (Bloch-like) are also possible; however, we will find that
even sharp interfaces result in extended conducting
states. Furthermore, we note that similar considerations apply to
step edges.

\begin{figure}
\includegraphics[width=3.3in]{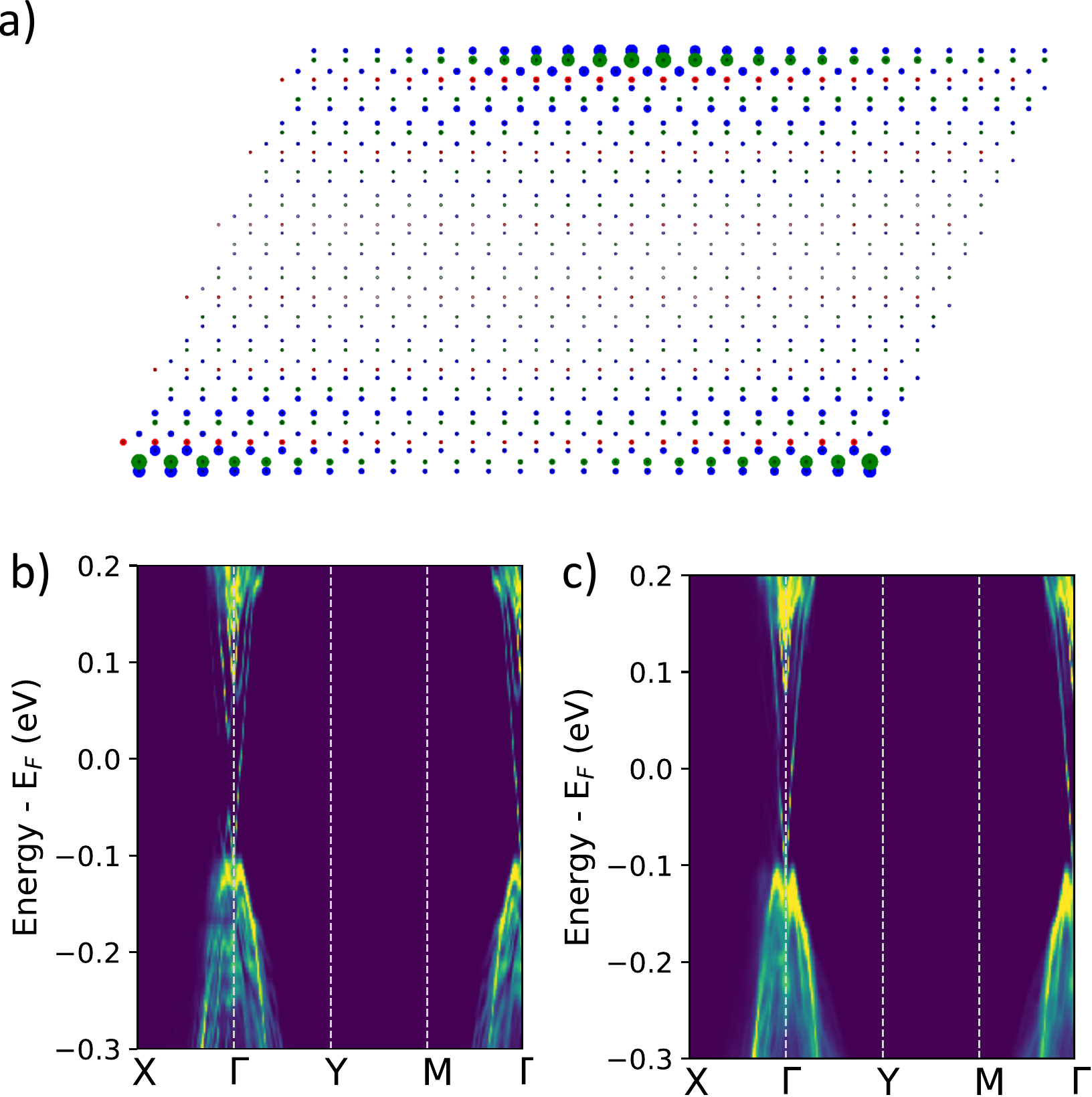}
\caption{\label{fig:domain} a) Real-space representation of $|\psi|^2$
  localized at domain wall in $24\times1\times5$ cell. Larger circles
  have more weight. Blue circles are Te, green are Bi, and red are
  Mn. b) Unfolded band structure in $20\times1\times5$ unit cell with
  10 unit cell domains. c) Average unfolded band structure (see text).}
\end{figure}

Using our model, we first study domains in a $24\times1\times5$ unit
cell, with surfaces in the $z$ direction, with two domains 12 unit
cells wide, and therefore two domain walls. As expected, we find a gap
closure at $k=\Gamma$, with four degenerate states. These states
correspond to the states localized at the two domain walls on each
surface, although at the degeneracy point they are all mixed
together. To make the plotting clearer, we move slightly away from
$\Gamma$, and consider two empty degenerate states at $k_x=0.05
\frac{2 \pi}{a} $.  In Fig.~\ref{fig:domain}a, we plot $|\psi|^2$ for
that pair of states, using larger circles to represent larger
magnitudes of the eigenvector. We find that as expected, the pair of
eigenvectors are surface states localized at the domain walls at $x=0$
and $x=0.5$ on the bottom and top surfaces. Even though we fix the
spins to reverse direction abruptly at the domain wall, we see that
the electronic states decay rather slowly perpendicular to the domain
wall, extending $\approx 10$ unit cells around the wall.

In Fig.~\ref{fig:domain}b, we consider the unfolded band structure for
a single example of a pair of domains, in a $20\times1\times5$ unit
cell. We see that there is a gap closure at $\Gamma$, and that the
band structure looks somewhat similar to the 2D topological surface state for
disordered spins (see Fig.~\ref{fig:temp}d), even though the metallic
edge channel is 1D. However, because we are only considering a single
pair of perfectly ordered and periodic domains, the unfolded topological
surface band has a variety of artifacts related to wavevectors of the
superlattice. In an experimental situation, we expect that there will
instead be domains of varying sizes. Therefore, in
Fig.~\ref{fig:domain}c, we average the unfolded surface band
structures of dozens of similar domains, with thicknesses of 4 to 10 unit
cells, in supercells of 8 to 20 unit cells. We see that we recover an average band structure that looks quite similar to the
2D topological surface state with disordered spins shown in
Fig.~\ref{fig:temp}d, even though every spin is perfectly aligned
along the $\pm z$ direction and the domain walls are sharp and
aligned. We expect that if we go
even further and include configurations with partially disordered
spins and domain walls in varying directions, the result will be band
structures that closely resemble the Dirac cone features we see at
high temperatures.

\section{Conclusions}

In conclusion, we generated a model to study the electronic structure
of large unit cells of the AFM topological insulator MnBi$_2$Te$_4$
with arbitrary spin configurations, which we have used to study three
types of topological phase transitions. First, we considered an
artificial transition driven by adjusting the magnitude of the SOC,
which proceeds via a bulk gap closure. Next, we considered a
topological transition driven by a temperature dependent magnetic
ordering. We find that as TRS is restored on average above $T_N$,
MnBi$_2$Te$_4$ goes from an AFM topological insulator with a surface
gap to a TRS-invariant $Z_2$ topological insulator with an associated
Dirac cone surface state, but with minimal change in the bulk
gap. Finally, we consider the electronic surface states associated
with AFM domain walls, which are 1D topological states. We find that
these states are strongly localized at the surface, but extend many
unit cells perpendicular to the domain walls, and that many 1D domain
walls can together resemble a Dirac cone-like topological surface
state on average.

Consistent with previous ideas, this work suggests that additional
sources of spin disorder beyond thermal fluctuations are necessary to
explain the gapless states observed experimentally at low
temperatures. We address this possibility more quantitatively, finding
that partially ordered surface spin configurations with broken TRS can
still result in a closed surface band gap. However, disorder fractions
above 50\% are necessary to fully close the surface gap. Possible
sources of disorder beyond thermal flucuations include quantum
fluctuations or chemical disorder. Alternatively, AFM domain walls can
produce electronic features that mimic the closed band gap seen above
$T_N$. Further experiments that quantify the local band gap of
MnBi$_2$Te$_4$ and that correlate the local gap with ARPES and
transport experiments may help clarify the topology and electronic
structure in this material.




%

\widetext
\clearpage

\begin{center}
\textbf{\large Supplementary materials: Topological surfaces states of MnBi$_2$Te$_4$ at finite temperatures and at domain walls}
\end{center}

\setcounter{equation}{0}
\setcounter{figure}{0}
\setcounter{table}{0}
\setcounter{page}{0}
\makeatletter


\renewcommand{\theequation}{S\arabic{equation}}
\renewcommand{\thefigure}{S\arabic{figure}}

Supplementary materials. Additional details on I. tight-binding model
evaluation II. the magnetic model III. the temperature dependent band
structure and topology.

\section{Tight-binding model evaluation}

\begin{figure}
\includegraphics[width=5.5in]{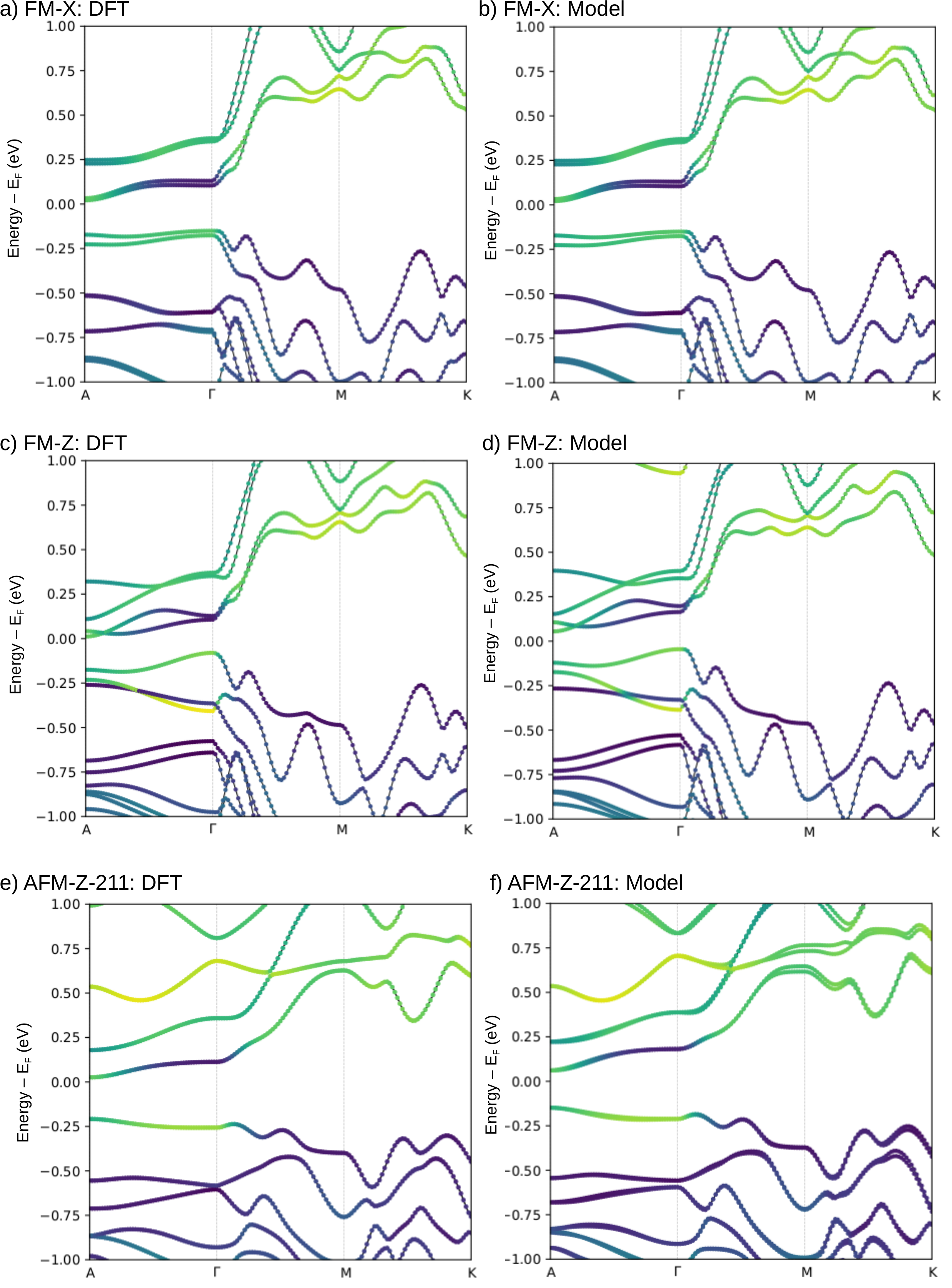}
\caption{\label{fig:compare} Comparison of DFT and model band structures, as in Fig. 1. a,c,e) DFT, b,c,d) Model. a-b) FM, spins in $x$-direction. c-d) FM, spins in $z$-direction, e-f) AFM, spins in $\pm z$ direction, alternating in-plane (not the ground state). }
\end{figure}

Fig.~\ref{fig:compare} shows several more bulk comparisons of direct
DFT-Wannier band structures with model band structures. Compare left
and right panels.

\section{Magnetic model}



\begin{figure}
\includegraphics[width=5.0in]{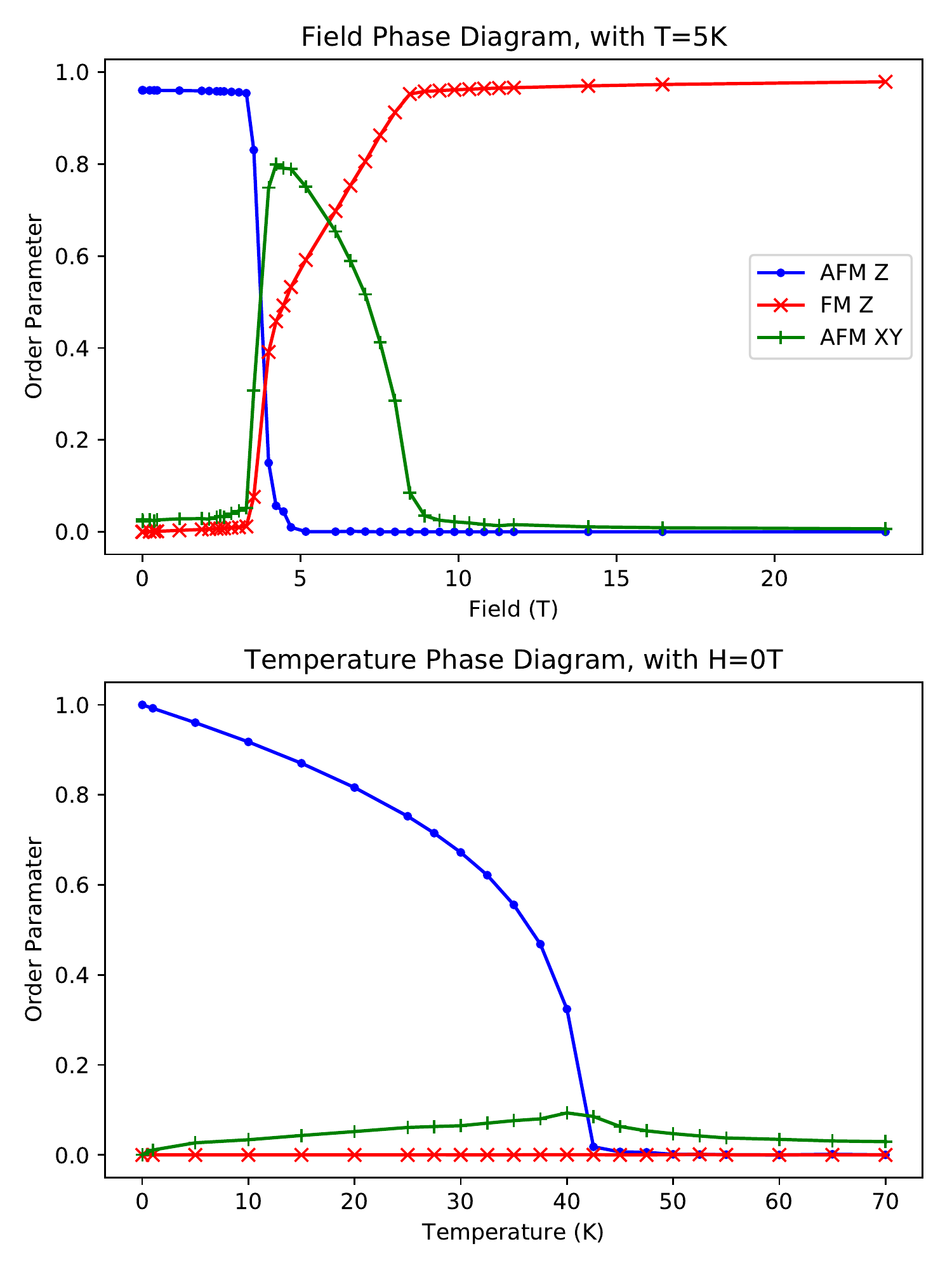}
\caption{\label{fig:magnetic_phase} Magnetic phase diagram as a function of field (top) and temperature (bottom). See text. Error bars are smaller than symbols.}
\end{figure}

We perform sampling in a $12\times12\times12$ unit cell for bulk
analysis of the model. To generate spin configurations for the
tight-binding model, we use a $3\times3\times5$ unit cell with the
model truncated along the $z$ direction to simulate a surface. In
Fig.~\ref{fig:magnetic_phase}, we show the magnetic phase diagram
under varying magnetic field at fixed temperature (top) and at varying
temperature and zero field (bottom). We consider the ground state AFM
$z$-direction spin configuration (blue), spins polarized FM along the
$z$-direction (red), and spins oriented AFM between layers, but
in-plane (green). We can observe a spin-flop transition in the top
panel under increasing field, followed by saturation at high field.

\begin{figure}
\includegraphics[width=5.0in]{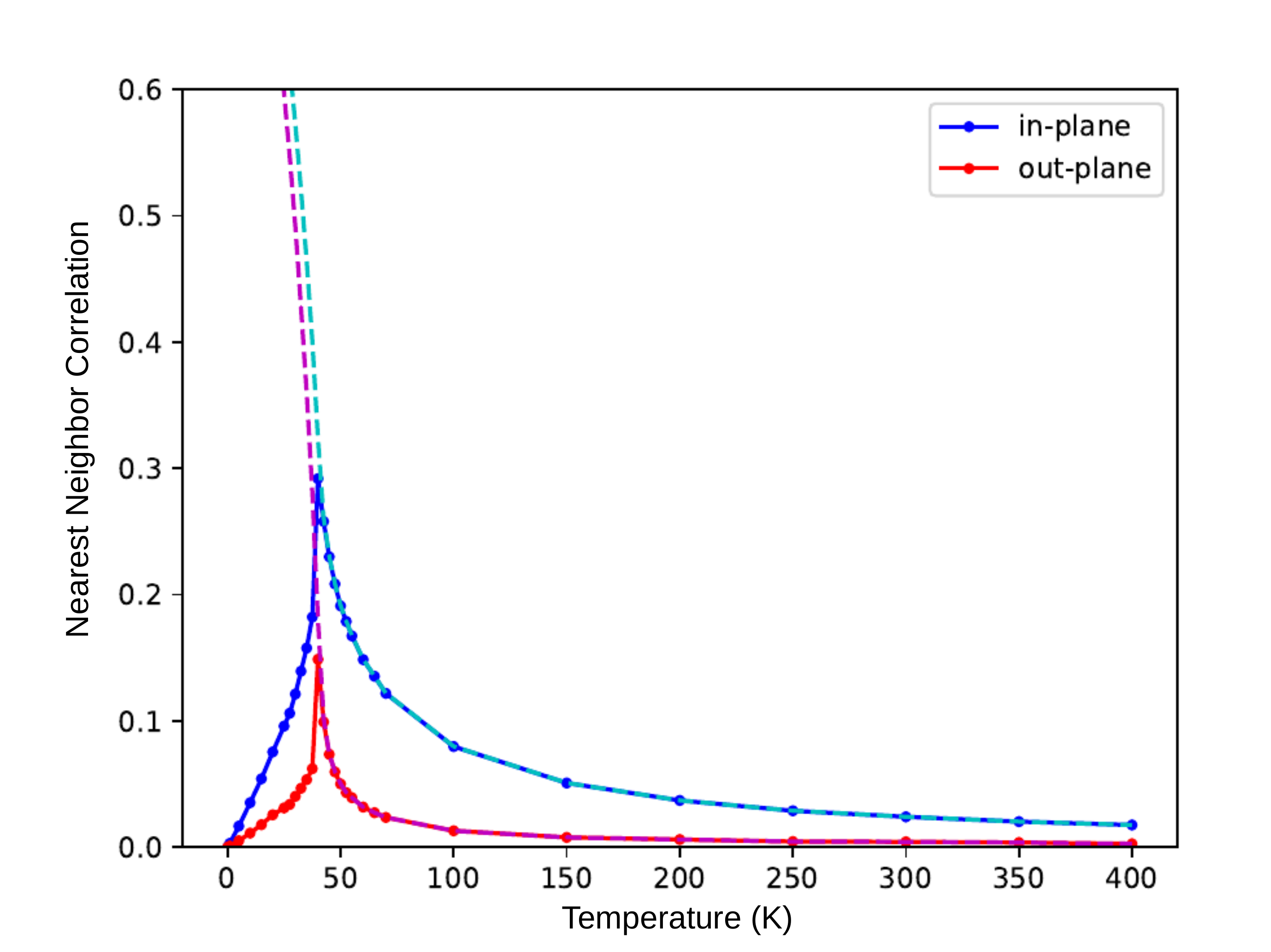}
\caption{\label{fig:corr} Magnetic nearest neighbor spin-spin
  correlation as a function of temperature at zero field. Blue line is
  in-plane correlation, red line is out-of-plane correlation. See text.}
\end{figure}

In Fig.~\ref{fig:corr}, we show the nearest neighbor in-plane (blue)
and out-of-plane (red) correlation as a function of temperature. Solid
lines show $\langle\vec{S_i}\cdot\vec{S_j}\rangle - \langle\vec{S_i}\rangle\cdot
\langle\vec{S_j}\rangle$, which goes to zero at low temperature, while dashed
lines show $\langle\vec{S_i} \cdot \vec{S_j}\rangle$, which goes to one below the
phase transition. In-plane correlations are much larger and remain significant to higher temperatures, which is
consistent with the shorter distances between Mn atoms in-plane and
the much larger magnetic interaction coefficients ($J_{ij}$) in-plane.

\section{Temperature Dependent Topology and Band Structures}

\begin{figure}
\includegraphics[width=5.0in]{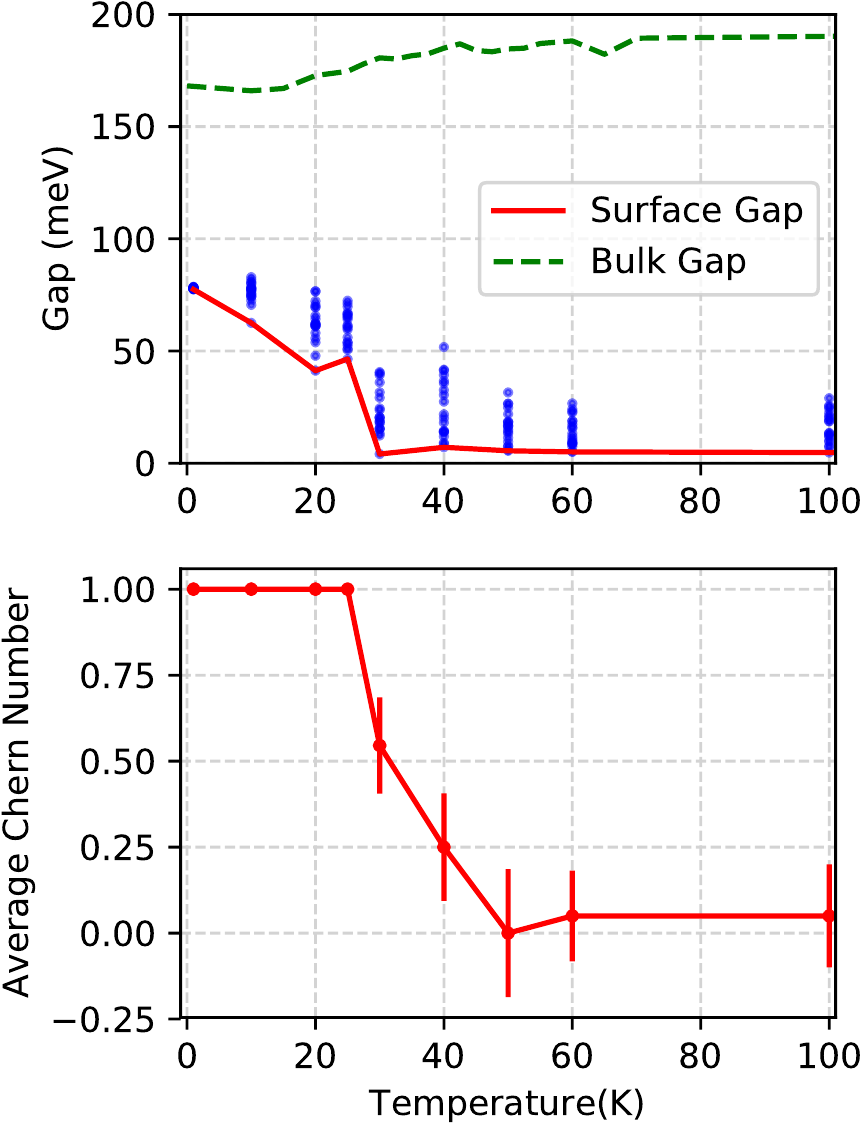}
\caption{\label{fig:temp2} Top: same as Fig. 2a. Bottom: Average Chern number as function of temperature. Error bars show 1 standard deviation statistical error.}
\end{figure}

Fig.~\ref{fig:temp2} (bottom) shows the average Chern number for the
$3\times3\times5$ with surfaces system studied in Fig. 2 in the main text, as a
function of temperature. While for any single magnetic snapshot, the
Chern number is always an integer, the average over many snapshots can
indicate how robust the topology is relative to magnetic
fluctuations. We expect that in a very large unit cell, instead of the
snapshots of a small periodic cell that we can calculate, the Chern
number would not fluctuate, but the local electronic structure and
local band gap would fluctuate and approach zero in regions close to
but below the N\'eel temperature.

\begin{figure}
\includegraphics[width=6.0in]{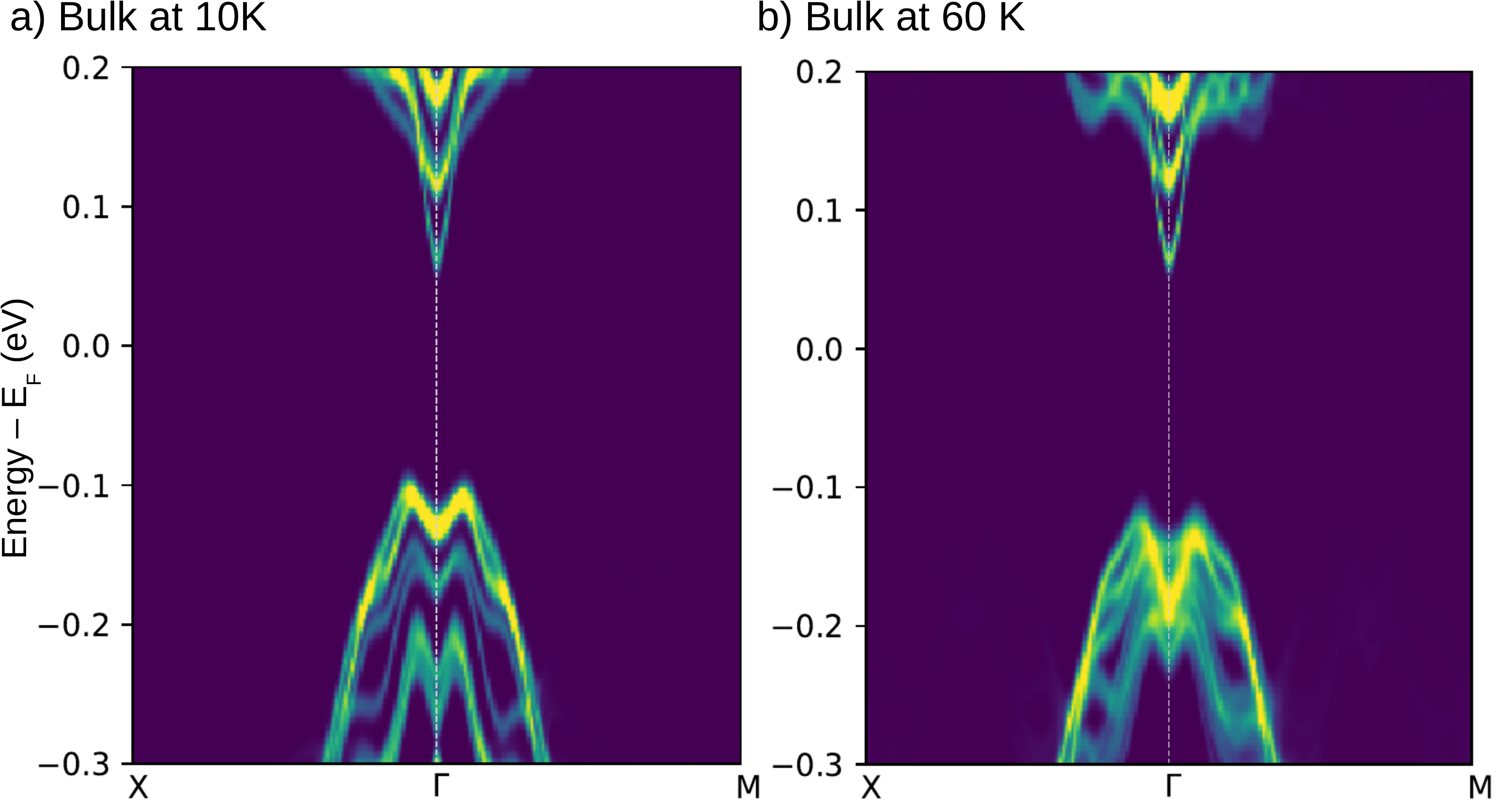}
\caption{\label{fig:bulk_bs} Bulk band structure in $3\times3\times6$
  unit cell, periodic in all three directions (no surface), unfolded
  to $1\times1\times6$ unit cell, for comparison with Fig. 2 b-d. a)
  10~K b) 60~K.}
\end{figure}

In Fig.~\ref{fig:bulk_bs}, we show bulk versions of the band
structures in Fig. 2 in the main text, in $3\times3\times6$ unit cells
unfolded to $1\times1\times6$ cells. Note the lack of states in the
gap, and the relatively small changes in electronic structure besides
averaging of spin-polarized bands above $T_N$.

%

\end{document}